# Optimizing Thermoelectric Power Factor by Means of a Potential Barrier

Neophytos Neophytou and Hans Kosina

Institute for Microelectronics, Technical University of Vienna, Austria

{neophytou | kosina}@iue.tuwien.ac.at

## Abstract

Large efforts in improving thermoelectric energy conversion are devoted to energy filtering by nanometer size potential barriers. In this work we perform an analysis and optimization of such barriers for improved energy filtering. We merge semiclassical with quantum mechanical simulations to capture tunneling and reflections due to the barrier, and analyze the influence of the width $W$, the height $V_b$, and the shape of the barrier, and the position of the Fermi level ($E_F$) above the band edge, $\eta_F$. We show that for an optimized design, ~40% improvement in the thermoelectric power factor can be achieved if the following conditions are met: $\eta_F$ is large; $V_b$–$E_F$ is somewhat higher but comparable to $k_B T$; and $W$ is large enough to suppress tunneling. Finally, we show that a smooth energy barrier is beneficial compared to a sharp (square) barrier for increasing the thermoelectric power factor.

# I. Introduction

The ability of a material to convert heat into electricity is measured by the dimensionless figure of merit $ZT=\sigma S^2 T/\kappa$, where $\sigma$ is the electrical conductivity, $S$ is the Seebeck coefficient, and $\kappa$ is the thermal conductivity. Traditional thermoelectric materials suffer from low efficiency with $ZT$ only around unity at the best case, attributed to the adverse interdependence of the electrical conductivity and the Seebeck coefficient via the carrier density, and to high thermal conductivities. Nanostructures and low-dimensional materials such as 1D nanowires (NWs) [1, 2], 2D thin-layer superlattices [3, 4, 5, 6], as well as materials with embedded nanostructuring [7, 8, 9, 10], however, have demonstrated significant enhancements in thermoelectric performance as a result of a large reduction in their lattice thermal conductivity.

Other than the reduction in the thermal conductivity, significant effort has been devoted in the use of nanostructures in improving the thermoelectric power factor as well. Current efforts target the modification of the electronic structure in order to achieve sharp features in the density of states function and strong energy dependence in the transmission coefficients. One of the initial ideas was discussed by Hicks and Dresselhaus [11], who suggested that low dimensional materials could provide improvements in the Seebeck coefficient and the power factor due to the sharp features in their density-of-states function. Mahan and Sofo [12] have predicted that infinitely large $ZT$ values could be achieved in zero-dimensional structures in the limit of zero lattice thermal conductivity. In our previous works, we showed that bandstructure engineering in low-dimensional channels through geometrical considerations such as confinement orientation, transport orientation and confinement length scale, can also provide power factor improvements [13, 14]. With regards to nanocomposite materials, theoretical work by Fomin e*t* al. in Ref. [15] have pointed out that large figure of merit values can be achieved in one-dimensional stacks of InAs/GaAs quantum dots upon the formation of mini-bands in the electronic structure of the material. Zianni in Ref. [16] has shown that large $ZT$ figures of merit can also be achieved in narrow nanowires whose diameter is modulated along the transport orientation resulting in strong energy dependence of their transmission coefficients. Approaches that utilize two-phase materials such as quantum dots embedded within a bulk medium [17] have also theoretically demonstrated large thermoelectric power factors. Two-phase materials based on heavily-doped nanocrystalline Si, in which case the two phases consist of the grains and the grain boundaries, demonstrated thermoelectric power factor values ~5× higher than bulk values [18]. In addition, it was shown that a simultaneous enhancement in the electrical conductivity and Seebeck coefficient can be achieved in such structures, resulting to very large power factors [17, 18]. In these approaches, the goal is to achieve improved energy filtering using one of the two phases of the material, which improves the Seebeck coefficient. The most usual strategy towards this, which has received much attention due to its experimental simplicity, is carrier energy filtering through nanoscale size potential energy barriers. Nanostructures based on superlattices and cross-plane transport, for example, have indeed demonstrated improvements in the Seebeck coefficient [19]. However, no significant improvements in the thermoelectric power factor $\sigma S^2$ of such structures were achieved to date, because in practice energy barriers also strongly reduce the electrical conductivity, although theoretical studies indicate that this should be possible [18, 20, 21].

In this work, we theoretically show that filtering by means of a nanometer scale potential barrier can indeed lead to power factor improvements. We lay down the features the barrier needs to have to achieve this goal. We merge semiclassical transport with quantum mechanical simulations to capture tunneling and reflections due to the barrier, and analyze the influence of the barrier's shape, width $W$, height $V_b$, and the influence of the Fermi level position in the region prior to the barrier, $\eta_F$ (as indicated in Fig. 1a). We show the power factor can indeed be improved by potential barriers whose height is of the order of $k_B T$ provided that the Fermi level is high into the region prior to the barrier such that high velocity carriers are utilized. Tunneling and quantum mechanical reflections from sharp barriers are detrimental to the power factor and should be avoided by the use of thicker and smooth barriers.

The paper is organized as follows: In section II we describe the simulation approach, in III we present and discuss the results, and in IV we conclude.

## II. Approach

The computational approach we employ is as follows: i) We semiclassically calculate the conductivity of the uniform material. ii) We introduce doping variation in the structure and solve Poisson's equation to obtain the potential barriers and depletion regions formed (Fig. 1a). The doping profiles we consider in the channels with the smooth barriers consist of the doping in the region prior to the barrier, and the doping in the barrier. We assume uniform doping in each region and for simplicity we ignore any dopant diffusion effects. In the case of the smooth barriers, the barrier is formed by the electrostatics of the different doping levels in the two regions. In the channels with the sharp barrier, the barrier region is undoped and the barrier height is set as an input parameter. iii) We use the non-equilibrium Green's function (NEGF) [22] formalism to calculate the quantum mechanical transmission from the barrier (Fig. 1b). iv) We compute the thermoelectric coefficients across the barrier.

To calculate the conductivity and Seebeck coefficient of the uniform material before the introduction of the barrier, we use the linearized Boltzmann equation. The electrical conductivity and Seebeck coefficient within the linearized Boltzmann theory are given by the following expressions [23]:

$$\sigma_0 = q_0^2 \int_{E_0}^{\infty} \sigma_0(E) dE, \qquad (1a)$$

with

$$\sigma_0(E) = \left(-\frac{\partial f_0}{\partial E}\right) \Xi(E), \qquad (1b)$$

and

$$S_0 = \frac{q_0 k_B}{\sigma_0} \int_{E_0}^{\infty} dE \left(-\frac{\partial f_0}{\partial E}\right) \Xi(E) \left(\frac{E - E_F}{k_B T}\right). \qquad (1c)$$

The transport distribution function $\Xi(E)$ is defined as [13, 23]:

$$\Xi(E) = N(E)\upsilon(E)^2 \tau(E) \qquad (2)$$

where $\upsilon(E)$ is the bandstructure velocity, $\tau(E)$ is the momentum relaxation time, and $N(E)$ is the 3D valence band density of states. We assume p-type bulk Si as the channel material because of the available experimental data, and the promise of Si as an effective thermoelectric material [1, 2, 24]. This should not impose, however, a loss in generality. We include phonon and impurity scattering, and calibrate our model to mobility measurements from literature in a large range of doping concentrations [25, 26, 27]. The calibration of p-type bulk Si mobility is described in detail in previous works [18, 28].

We employ the NEFG formalism to compute the quantum mechanical transmission function $T(E)$ across the barrier [22]. In this method, the Green's function is defined as:

$$G(E) = [EI - H - \Sigma_1 - \Sigma_2]^{-1} \qquad (3)$$

where $I$ is the identity matrix, $H$ is the device Hamiltonian and $\Sigma_1$ and $\Sigma_1$ are the self-energies for the contacts. The device is considered to be the region as shown in Fig. 1a and Fig. 1b, which includes flat potential regions in the left and right sides, and the barrier in the middle. The contacts are considered to be semi-infinite, connected to the far left and far right sides of the device domain. We consider a 1D problem for the calculation of the transmission coefficient and an effective mass description for the Hamiltonian. In this case the self-energies are given analytically by $\Sigma_{1/2} = -t_0 e^{-ik_{1/2}\alpha}$, where $k_{1/2}$ is the wavevector and $\alpha$ is the lattice unit cell size, in this case the discretization length of the simulation domain. Using the finite difference method, the on-site elements of the Hamiltonian are given by $h_0 = 2t_0 + U$ and the off-site elements by $h_1 = -t_0$, where $t_0 = \hbar^2/(2m^*\alpha^2)$ is the coupling constant, and $U$ is potential. For the effective mass we use $m^* = 0.49 m_0$, which corresponds to the Si heavy-hole mass [27].

The transmission probability versus energy is then calculated as $T(E) = trace(\Gamma_1 G \Gamma_2 G^+)$ with $\Gamma_{1/2} = i(\Sigma_{1/2} - \Sigma_{1/2}^+)$.

The electrical conductivity across the barrier is calculated by the product of the conductivity in the uniform material multiplied by the quantum mechanical transmission function *T(E)* obtained using NEGF:

$$\sigma_b(E) = \sigma_0(E) T(E) \qquad (4a)$$

The Seebeck coefficient across the barrier is calculated in a similar manner as:

$$S_b = \frac{q_0 k_B}{\sigma_b} \int_{E_0}^{\infty} dE \left(-\frac{\partial f_0}{\partial E}\right) \Xi(E) T(E) \left(\frac{E - E_F}{k_B T}\right) \qquad (4b)$$

Figures 1c and 1d show the energy dependence of the conductivity $\sigma(E)$ and of the Seebeck coefficient $S(E)$ for different example cases. The uniform bulk material coefficients are shown in blue, the coefficients across the smooth barrier in black-solid, and those across the square barrier in black-dashed lines. The conductivity is reduced compared to bulk after the introduction of the barriers. Interestingly, the conductivity is reduced more in the case of the square barriers compared to the smooth ones. The Seebeck coefficient, on the other hand, is increased with the introduction of barriers (Fig. 1d). It is increased slightly more in the case of square barriers, following the usual inverse trend compared to conductivity.

## III. Results and Discussion

*Influence of the barrier height $V_b$:* We first investigate the effect of the barrier height $V_b$ on the thermoelectric coefficients. For this we place the Fermi level at $\eta_F =$ 0.84eV above the band edge, which corresponds to a carrier density of $p=10^{20}$cm$^{-3}$. By changing the doping concentration in a small part of the channel (*W*), we introduce a built-in barrier. We set the width of this part of the channel to *W*=2nm. Different barrier heights can be obtained by changing the concentration. The actual barrier height is obtained by solving the Poisson equation. We increase the height from $V_b$=75meV to $V_b$=192meV. Figure 2a shows the reduction of the electrical conductivity compared to the bulk value as the barrier is increased. The conductivity drops by ~6× as the barrier is increased. The conductivity of the channel with the smooth barrier is always higher than that with a square barrier of the same width $W$=2nm. On the other hand, the Seebeck coefficient in Fig. 2b is increased as expected, since barriers enhance the effect of energy filtering. A factor of ~1.7× increase is observed, with the square barrier providing slightly higher Seebeck coefficient compared to the smooth barrier (except for very high $V_b$). The power factor is shown in Fig. 2c. As expected, a maximum is observed for the power factor, which is ~15% higher than the power factor of the bulk material. The maximum appears for barrier heights ~30meV above the Fermi level for the smooth barrier. This optimum $V_b$ value, close to $k_BT$ (or somewhat higher), is also in agreement with other theoretical studies [18, 21]. In the case of the square barriers we find that the optimum appears for lower barrier heights (but as we shall explain below this can depend on the width and effective mass of the barrier as well). In the case of the smooth barrier, the maximum in the power factor is higher than that of the square barrier, and is retained at high levels for a large range of barrier heights. The advantage of the smooth barrier over the square barrier is due to its higher conductivity. At the position of the optimum power factor its conductivity is ~20% higher, whereas its Seebeck coefficient is only slightly smaller (~5%) as shown in Fig. 1c and 1d.

*Influence of the Fermi level position, $\eta_F$:* We next investigate the influence of the distance $\eta_F$ of the Fermi level to the band edge in the uniform material prior to the barrier. In this case, we vary the doping density in the channel to alter $\eta_F$. We then adjust the concentration in the narrow ($W$=2nm) low doped region such that the barrier is always 30meV higher than the Fermi level, i.e. $V_b$-$E_F$ =30meV. This is the energy distance that maximizes the power factor as shown earlier in Fig. 2c, and is kept fixed. The simulation results for the conductivity, Seebeck coefficient, and power factor versus $\eta_F$ are shown in Fig. 3a, 3b, and 3c, respectively. The conductivity is increased in all cases, the bulk (blue), the smooth barrier (black-solid), and the square barrier (black-dashed), as the Fermi level raises. An overall increase of ~10×, ~6× and ~5×, respectively, is observed as the Fermi level raises from -10meV to 110meV above the band edge. These values correspond to carrier densities of $1.1 \times 10^{19} cm^{-3}$ to $1.6 \times 10^{20} cm^{-3}$. The conductivity increase can be attributed to two reasons. The first is that the higher the Fermi level is, the higher the carrier velocities are, which increases conductivity. The second is that at such high doping concentrations the channel conductivity is limited by impurity scattering. The higher the carrier density is, the stronger the impurity screening by free carriers becomes, which weakens the Coulomb potential. These effects reduce the impurity scattering rate, significantly raise the mean-free-path for scattering and thus the electrical conductivity. On the other hand, the Seebeck coefficient shown in Fig. 3b is reduced, following the inverse trend compared to the electrical conductivity. However, the reduction is much smaller than the increase in the conductivity. A reduction of ~60%, ~40%, and ~35% is observed for the cases of the bulk, smooth barrier, and square barrier channels, respectively. Due to the much larger increase in the electrical conductivity compared to the reduction in Seebeck coefficient, the power factor $\sigma S^2$ is improved by raising the Fermi level. The importance of the large $\eta_F$ is also verified by other theoretical studies [20], but also experimentally in Ref. [18], where we pointed out that large thermoelectric power factors can be achieved in nanocrystalline channels which meet this condition. The channel with the smooth barrier has a higher power factor than the channel with the square barrier due to its higher conductivity. The channel with the square barrier has a slight advantage over bulk at very high Fermi levels. Interestingly, although the power factor of the bulk uniform material saturates and begins to slightly decrease after $\eta_F$ ~80meV, the power factor in the channels that include the smooth barrier continues to increase in the range up to $\eta_F$ ~110meV that we examine.

To raise the Fermi level so high in that region, the doping can be as high as $1.6 \times 10^{20} cm^{-3}$ in the regions prior to the barriers, but significantly lower in the barrier regions. Several works in the literature indicate that impurity energy bands could be introduced in the Si bandstructure at such high doping concentrations (in this case in the highly doped region prior to the barrier) [29, 30, 31, 32]. As shown by Yamashita *et al.*, for p-type Si, bandstructure effects could appear for carrier concentrations above $p=3\times10^{19}$ cm$^{-3}$ [29]. Studies indicate that these bands could improve the Seebeck coefficient and power factor even further [30, 32], although impurity bands might make it difficult for the Fermi level to rise to the levels suggested above. In this work, however, for simplicity, we do not consider such effects. We focus on optimizing the barrier characteristics for efficient energy filtering, and any additional bandstructure effects are out of this scope, although we do not expect that our qualitative conclusions will be altered in the presence of impurity bands or different bandstructures in general. It is not clear if impurity bands will be present in the two-phase materials that we examine since their formation requires extended uniformly doped channels, which will be determined by the length of the doped region we consider. The formation of impurity bands also depends on the choice of the channel material. We note that the choice of p-type Si parameters in this work is only for the calibration of the simulations to available mobility measurements. Our findings and design guidelines, however, would be generally applicable for different materials as well, for which impurity bands might, or might not occur.

*Influence of the barrier width, W:* The third design parameter that we examine is the width of the barrier, $W$. Again, in this case we keep $\eta_F$ ~84meV and $V_b$ =115meV, the parameters which provided the maximum power factor in Fig. 1c. Figure 4 shows the change in the thermoelectric coefficients as the width of the barrier changes from $W$=1nm to $W$=10nm. In Fig. 4a the electrical conductivity shows an initial decrease with increasing width, attributed to the suppression of tunneling, and then remains almost constant. Other than for the very thin, transparent barrier with $W$=1nm, the conductivity in the channel with the smooth barrier is by ~25% larger than that of the square barrier. The reason is that the quantum mechanical reflections due to a sharp barrier are stronger than due to a smooth barrier. This is clearly indicated in the energy dependence of the conductivity of the $W$=2nm barrier in Fig. 1c, which shows that $\sigma(E)$ is higher in the case of the smooth barrier. In Fig. 5a the same is shown for the wider barriers with $W$=10nm. The blue line shows the conductivity as a function of energy for the bulk material. As indicated by the black-solid line, the conductivity for the material with a smooth barrier is suppressed at low energies and bounded by that of the bulk channel at higher energies.

The black-dashed line shows the same quantity for the channel with the square barrier. The quantum mechanical oscillations reduce the conductivity below that of the channel with the smooth barrier. On the other hand, the Seebeck coefficient is higher in the channel with the square barrier, as shown in Fig. 5b for the $W$=10nm case. Figure 4b shows the dependence of the Seebeck coefficient on the barrier width. For very thin barriers, where tunneling carriers contribute significantly to the channel conductivity, the Seebeck coefficient is lower. The Seebeck coefficient increases gradually by ~20% (smooth barrier) and ~55% (square barrier) as the width increases to $W$~4nm. For the wider barriers, tunneling is completely suppressed (as shown in Fig. 5a the conductivity for energies below $V_b$ is negligible for the channel with a $W$=10nm thick barrier). We note that although the tunneling current is suppressed, the electrical conductivity does not suffer significantly because the transmission above the barrier is larger for wider barriers (compare the peak of $\sigma(E)$ in Fig. 5a with that of Fig. 1c). For widths above $W$=4nm the Seebeck coefficient saturates for channels that include either barrier. The square barrier provides a ~10% larger Seebeck coefficient. The power factor versus width is shown in Fig. 4c. A significant increase compared to the bulk power factor is achieved in the channels with barriers. The power factor saturates beyond $W$~5nm at values ~25% and ~40% higher than the bulk power factor in the channels with square and smooth barriers, respectively. We note that a similar dependence on width is observed if the effective mass of the material is increased. In that case, the power factor improvement for both barriers shown in Fig. 2 and Fig. 3 would be higher compared to what we show now, but the width saturation behavior in Fig. 4 will occur at thinner barriers (e.g. a larger benefit of the square barrier over the bulk channel would be evident in the study case in Fig. 3 if a wider barrier was assumed).

It is interesting to note that the power factor improvements are resulting to a large degree from conductivity improvements. In all three case studies we performed, (i.e. variations of $V_b$, $\eta_F$, and $W$), the highest power factor is observed when the electrical conductivity is enhanced. For example, the conductivity of the channel with a smooth barrier is always higher than that of the square barrier channel. Although the Seebeck coefficient is in most cases lower, the power factor is always higher in channels with smooth barriers. In addition, the influence of the design parameters we examine (with the exception of the very thin, transparent $W$=1nm wide barriers) are larger on the conductivity than on the Seebeck coefficient, which is affected much less. Upon $V_b$ increase, the conductivity is reduced by ~5-6×, whereas the Seebeck improves only by ~2×. Upon increase in $\eta_F$, the conductivity is increased by ~5-6×, whereas the Seebeck decreases only by ~40%. Interestingly, although the barriers are introduced in order to improve filtering and thus to improve the Seebeck coefficient, improvements to the power factor can only be achieved when the conductivity in the region prior to the barrier is high, which improves the overall conductivity of the channel with the barrier. Additionally, the highest power factors are achieved in the channels with smooth barriers that have the largest conductivity (despite the lower Seebeck coefficient). A similar conclusion regarding the importance of the electrical conductivity was also reached in the cases of uniform low-dimensional Si channels [13, 14]. In those works, it was pointed out that changes in the power factor due to geometry variations, or low-dimensionality, originate mostly from the electrical conductivity which was strongly affected, rather than the Seebeck coefficient which was affected less.

We finally note that the ~10% advantage of the smooth barrier over the sharp barrier is only relevant to the thermoelectric power factor and not necessarily the *ZT* figure of merit. Sharp potential barriers are usually formed in heterostructures consisting of two different materials. In these structures the thermal conductivity is reduced drastically due to enhanced phonon-boundary scattering at the interfaces of the two materials [3]. Indeed, studies on the thermal conductivity of one-dimensional composite structures with sharp geometrical features, i.e. in diameter modulated SiC nanowires [33], or quantum-dot superlattices [34], have indicated that remarkably low thermal conductivities can be achieved, which could provide very large *ZT* values. Smooth barriers, on the other hand are usually achieved with doping variation along the channel, or modulation doping techniques. In these structures the thermal conductivity reduction due to dopant impurities is weaker. The *ZT* figure of merit, therefore, could be larger in the sharp barrier geometry (which has very low thermal conductivity) compared to the smooth one (which could have larger thermal conductivity).

## IV. Conclusions

In summary, we have investigated thermoelectric carrier energy filtering from a potential barrier at room temperature. We show that a ~40% improvement in the thermoelectric power factor can theoretically be achieved. For maximizing energy filtering: i) In the region prior to the barrier the Fermi level needs to be pushed high such that carriers have higher velocity and scattering by ionized dopant impurities is weaker, ii) The barrier height for maximum energy filtering is $V_b$-$E_F \sim k_B T$ (or somewhat larger) and iii) The width of the barrier $W$ is large enough to suppress tunneling, which turns out to be detrimental to the Seebeck coefficient. Our analysis on smooth and square-shaped barriers showed that the former are more beneficial compared to the latter in increasing the thermoelectric power factor. This is due to the fact that in smooth barrier channels, conductivity does not suffer from the strong quantum mechanical reflections that occur in square-like barrier channels.

*Acknowledgement:* The work leading to these results has received funding from the European Community's Seventh Framework Programme under grant agreement no. FP7-263306.

# References

[1] A.I. Boukai, Y. Bunimovich, J. T.-Kheli, J.-K. Yu, W. A. Goddard III, and J. R. Heath, *Nature*, 451, 168-171, 2008.

[2] A. I. Hochbaum, R. Chen, R. D. Delgado, W. Liang, E. C. Garnett, M. Najarian, A. Majumdar, and P. Yang, *Nature*, 451, 163-168, 2008.

[3] R. Venkatasubramanian, E. Siivola, T. Colpitts, and B. O' Quinn, *Nature*, 413, 597-602, 2001.

[4] D. Li, Y. Wu, R. Fang, P. Yang, and A. Majumdar, *Appl. Phys. Lett.*, vol. 83, no. 15, pp. 3186–3188, 2003.

[5] W. Kim, S. L. Singer, A. Majumdar, D. Vashaee, Z. Bian, A. Shakouri, G. Zeng, J. E. Bowers, J. M. O. Zide, and A. C. Gossard, *Appl. Phys. Lett.*, 88, 242107, 2006.

[6] G. Zeng, J.E Bowers, J.M.O. Zide, A.C. Gossard, W. Kim, S. Singer, A. Majumdar, R. Singh, Z. Bian, Y. Zhang, A. Shakouri, *Appl. Phys. Lett.*, 88, 113502, 2006.

[7] J. Tang, H.-T. Wang, D. H. Lee, M. Fardy, Z. Huo, T. P. Russell, and P. Yang, *Nano Lett.*, 10, 10, 4279-4283, 2010.

[8] C. J. Vineis, A. Shakouri, A. Majumdar, and M. C. Kanatzidis, *Adv. Mater.*, 22, 3970-3980, 2010.

[9] L.D. Zhao, S. H. Lo, J. Q. He, L. Hao, K. Biswas, J. Androulakis, C. I. Wu, T. P. Hogan, D. Y. Chung, V. P. Dravid, and M. G. Kanatzidis, *J. of the American Chemical Society*, 2011,133, 20476-20487.

[10] Y. He, D. Donadio, and G. Galli, *Nano Lett.*, DOI: 10.1021/nl201359q, 2011.

[11] L.D. Hicks, M. S. Dresselhaus, *Phys. Rev. B*, *47*, 16631, 1993.

[12] G. D. Mahan, J. O. Sofo, *Proc. Natl. Acad. Sci. USA*, *93*, 7436-7439, 1996.

[13] N. Neophytou, H. Kosina, *Phys. Rev. B*, *83*, 245305, 2011.

[14] N. Neophytou, H. Kosina, *J. Appl. Phys.*, 112, 024305, 2012.

[15] V. M. Fomin and P Kratzer, *Phys. Rev. B*, 82, 045318, 2010.

[16] X. Zianni, *Appl. Phys. Lett.*, 97, 233106, 2010.

[17] J. Zhou and R. Yang, *J. Appl. Phys.*, 110, 084317, 2011.

[18] N. Neophytou, X. Zianni, H. Kosina, S. Frabboni, B. Lorenzi, and D. Narducci, Nanotechnology, 24, 205402, 2013.

[19] G. Zeng, J. M. O. Zide, W. Kim, J. E. Bowers, A. C. Gossard, Z. Bian, Y. Zhang, A. Shakouri, S. L. Singer, and A. Majumdar, *J. Appl. Phys.*, 101, 034502, 2007.

[20] D. Vashaee, and A. Shakouri, *Phys. Rev. Lett.*, *92*, 106103, 2004.

[21] R. Kim and M. Lundstrom, *J. Appl. Phys.*, 110, 034511, 2011.

[22] S. Datta, *Electronic Transport in Mesoscopic Systems*, Cambridge Univ. Press, Cambridge MA, 1997.

[23] T. J. Scheidemantel, C. A.-Draxl, T. Thonhauser, J. V. Badding, and J. O. Sofo, *Phys. Rev. B*, vol. 68, p. 125210, 2003.

[24] K. Nielsch, J. Bachmann, J. Kimling, and H. Boettner, *Adv. Energy Mat.*, 1, 713, 2011.

[25] C. Jacoboni and L. Reggiani, *Rev. Mod. Phys.*, vol. 55, 645, 1983.

[26] G. Masetti, M. Severi, and S. Solmi, *IEEE Trans. Electr. Dev.*, 30, 764, 1983.

[27] http://www.ioffe.ru/SVA/, "Physical Properties of Semiconductors".

[28] N. Neophytou, X. Zianni, M. Ferri, A. Roncaglia, G. F. Cerofolini, D. Narducci, *J. Electr. Mat.*, 42, 2393-2401, 2013.

[29] O. Yamashita and N. Sadatomi, *Jap. J. Appl. Phys.*, 38, pp. 6394-6400, 1999.

[30] H. Ikeda and F. Salleh, *Appl. Phys. Lett.*, 96, 012106, 2010.

[31] F. Salleh and H. Ikeda, *J. Electronic Mat.*, 40, 903, 2011.

[32] A. Popescu and L. M. Woods, *Appl. Phys. Lett.*, 97, 052102, 2010.

[33] K. Termentzidis, T. Barreteau, Y. Ni, S. Merabia, X. Zianni, Y. Chalopin, P. Chantrenne, and S. Voltz, *Phys. Rev. B*, 87, 125410, 2013.

[34] D. L. Nika, E. P. Pokatilov, A. A. Balandin, V. M. Fomin, A. Rastelli, and O. G. Schmidt, *Phys. Rev. B*, 84, 165415, 2011.

Figure 1:

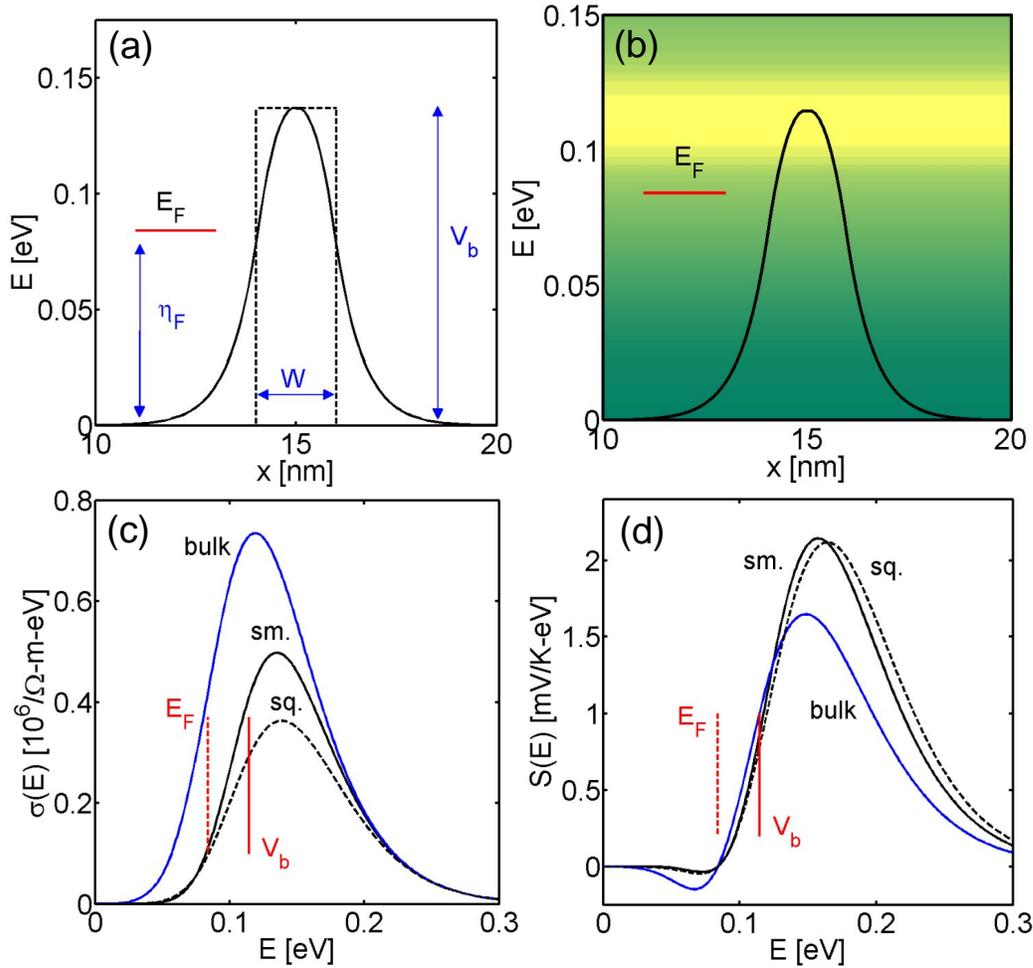

Figure 1 caption:

(a) The barrier shapes used in the simulations: smooth and square. The three parameters investigated are indicated as $\eta_F$, $V_b$, and $W$. In the case of the smooth barrier, $W$ is defined as the width of the lightly doped region that creates the barrier. (b) Example of the transmission function through a smooth barrier weighted by the derivative of the Fermi distribution (Eq. 1a), as a function of energy and space. (c) The energy dependence of the conductivity. (d) The energy dependence of the Seebeck coefficient. (c) and (d) show results for the uniform bulk material (blue), a smooth barrier of $W=2$nm (black-solid), and a square barrier of $W=2$nm (black-dashed). The positions of the Fermi level and the barrier height are indicated at $E_F=84$meV and $V_b=115$meV, respectively.

Figure 2:

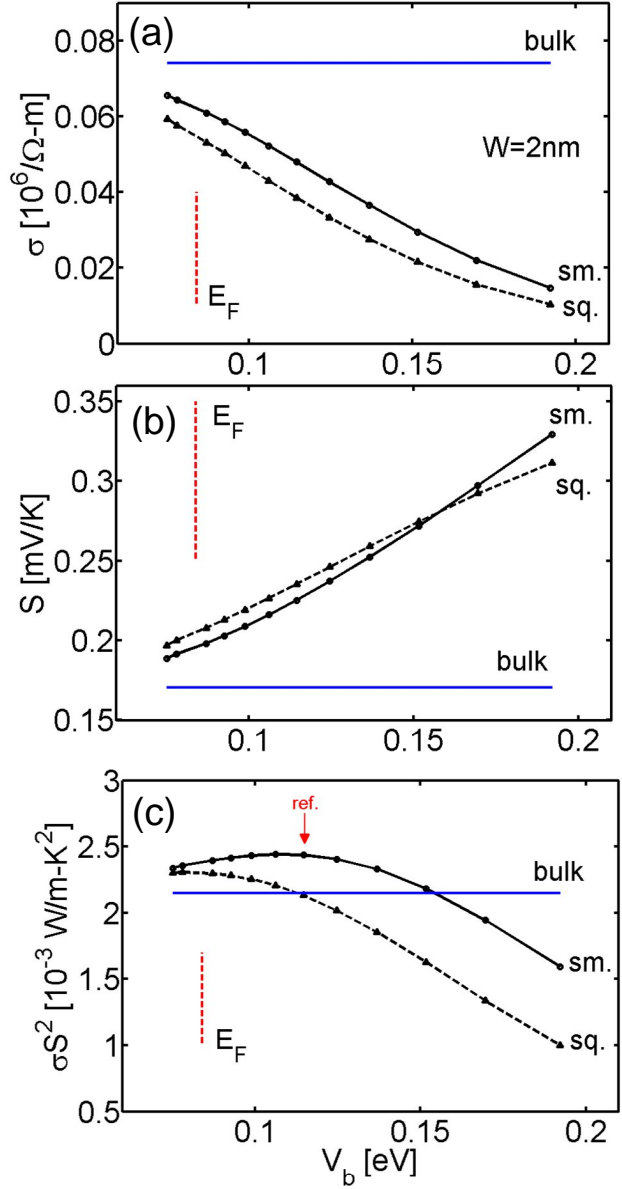

## Figure 2 caption:

Influence of the barrier height $V_b$ on the thermoelectric coefficients: (a) Electrical conductivity, (b) Seebeck coefficient, and (c) power factor. The width of the barrier is $W$=2nm, and $\eta_F$ =84meV (the position of the Fermi level is indicated by the red-dashed line). Blue-solid line: uniform bulk material. Black-solid line: Material with a smooth barrier. Black-dashed line: Material with a sharp square barrier. The red arrow indicates the reference device for comparisons in Fig. 3 and Fig. 4.

Figure 3:

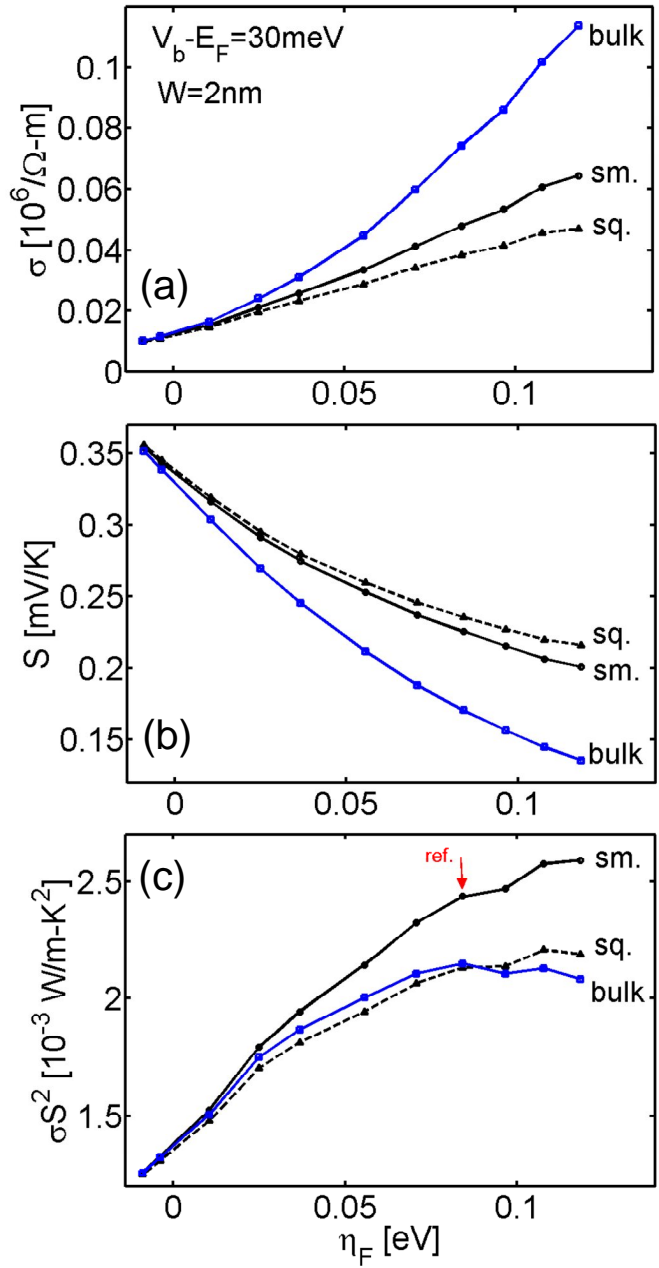

## Figure 3 caption:

Influence of the position of the Fermi level $\eta_F$ on the thermoelectric coefficients: (a) Electrical conductivity, (b) Seebeck coefficient, and (c) power factor. The width of the barrier is $W=2$nm, and the height $V_b=115$meV. Blue-solid line: uniform bulk material. Black-solid line: Material with a smooth barrier. Black-dashed line: Material with a sharp square barrier. The red arrow indicates the reference device as in Fig. 2.

Figure 4:

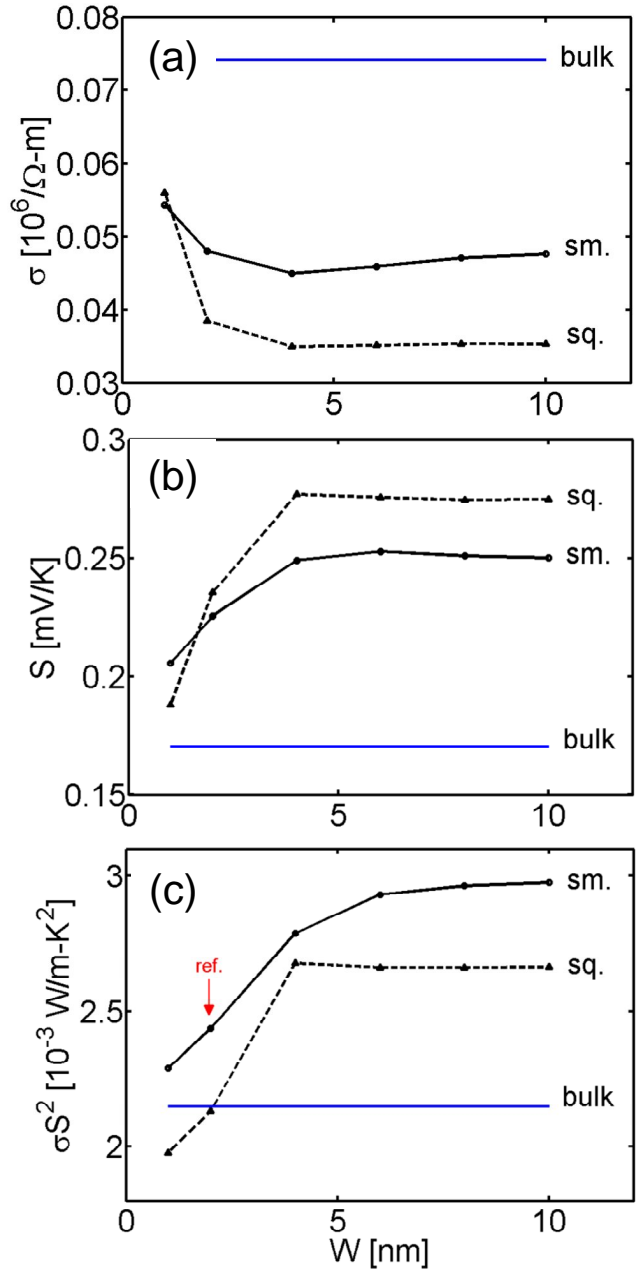

## Figure 4 caption:

Influence of the barrier width $W$ on the thermoelectric coefficients: (a) Electrical conductivity, (b) Seebeck coefficient, and (c) power factor. The barrier height is $V_b$=115meV and $\eta_F$ =84meV. Blue-solid line: uniform bulk material. Black-solid line: Material with a smooth barrier. Black-dashed line: Material with a square barrier. The red arrow indicates the reference device as in Fig. 2.

Figure 5:

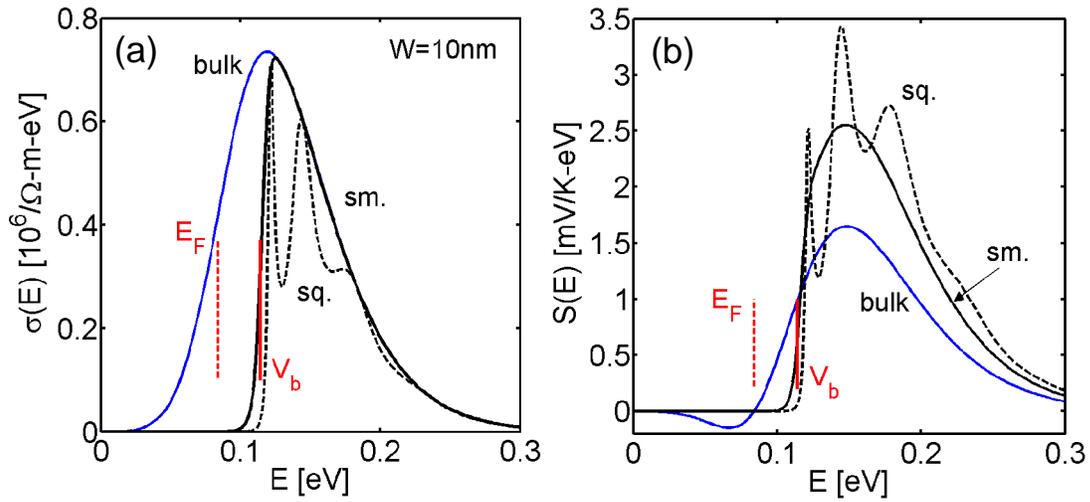

## Figure 5 caption:

Energy dependence of the electrical conductivity (a) and Seebeck coefficient (b) for channels of width $W$=10nm. Blue-solid line: uniform bulk material. Black-solid line: Material with a smooth barrier. Black-dashed line: Material with a square barrier. The positions of the Fermi level and the barrier height are indicated.

19